\documentclass[twocolumn,aps,prl,superscriptaddress]{revtex4-2}

\usepackage{amsmath}
\usepackage{graphicx}
\usepackage{MnSymbol}
\usepackage{parskip}

\begin{document}

\title{Rubber Wear: History, Mechanisms, and Perspectives}

\author{R. Xu}
\affiliation{State Key Laboratory of Solid Lubrication, Lanzhou Institute of Chemical Physics, Chinese Academy of Sciences, 730000 Lanzhou, China}
\affiliation{Shandong Linglong Tire Co., Ltd, 265400 Zhaoyuan, China}
\affiliation{Peter Gr\"unberg Institute (PGI-1), Forschungszentrum J\"ulich, 52425, J\"ulich, Germany}
\affiliation{MultiscaleConsulting, Wolfshovener str. 2, 52428 J\"ulich, Germany}
\author{W. Sheng}
\affiliation{State Key Laboratory of Solid Lubrication, Lanzhou Institute of Chemical Physics, Chinese Academy of Sciences, 730000 Lanzhou, China}
\author{F. Zhou}
\affiliation{State Key Laboratory of Solid Lubrication, Lanzhou Institute of Chemical Physics, Chinese Academy of Sciences, 730000 Lanzhou, China}
\author{B.N.J. Persson}
\affiliation{State Key Laboratory of Solid Lubrication, Lanzhou Institute of Chemical Physics, Chinese Academy of Sciences, 730000 Lanzhou, China}
\affiliation{Shandong Linglong Tire Co., Ltd, 265400 Zhaoyuan, China}
\affiliation{Peter Gr\"unberg Institute (PGI-1), Forschungszentrum J\"ulich, 52425, J\"ulich, Germany}
\affiliation{MultiscaleConsulting, Wolfshovener str. 2, 52428 J\"ulich, Germany}

\begin{abstract}
This paper presents a comprehensive review of wear mechanisms, with a primary focus on rubber wear under sliding conditions. Beginning with classical wear theories, including the Archard and Rabinowicz models, we analyze their applicability to both metals and elastomers and discuss extensions relevant to elastic contact and multiscale surface roughness. Various experimental studies on rubber, such as abrasion by sharp tools, erosion by particle impact, and wear during sliding on rough substrates, are reviewed and interpreted. The effects of environmental factors, such as oxygen and lubrication, are also discussed. In addition, we review a recently proposed wear model based on fatigue crack growth within asperity contact regions, which accounts for energy dissipation and multiscale interactions. This model explains the wide variability in the wear coefficient and predicts wear rates consistent with experimental observations across a broad range of conditions. It also explains the formation mechanism and provides the size distribution of rubber wear particles, from micrometer-scale up to severe cut-chip-chunk (CCC) wear. The results have implications for tire wear and the environmental impact of microscale wear debris (microplastics).
\end{abstract}

\maketitle

\setcounter{page}{1}
\pagenumbering{arabic}




{\bf Corresponding author:} B.N.J. Persson, email: b.persson@fz-juelich.de
\vskip 0.3cm

{\bf 1 Introduction}

Wear is defined as the damage to a solid surface, generally involving the progressive loss of material, due to relative motion between that surface and a contacting substance or substances. This does not only involve solid-solid contact, as the term ``substance'' could also refer to a liquid, as in cavitation erosion.

Although often viewed as a technical issue, wear has posed challenges to humans for thousands of years. Even back in the Stone Age, hunters needed to consider the importance of maintaining the sharpness of their stone axes or knives. Nowadays, wear is not only a major reason for the failure of mechanical devices, but also a contributor to air pollution. In recent decades, environmental concerns related to wear have received increasing attention. Particles released from wear processes, such as those from tires or brakes, pose significant health risks. Particles with aerodynamic diameters greater than $\sim 10 \ {\rm \mu m}$ are typically deposited in the nose or throat and cannot penetrate the lower tissues of the respiratory tract. However, smaller particles can enter the lungs and translocate into vital organs due to their size, causing significant human health consequences. The degree of toxicity becomes greater for smaller particles.

The wear resistance of a material depends to a large extent on its conditions of use \cite{Blau}. Nevertheless, it is frequently mischaracterized as an inherent property of the material. In fact, any judgment regarding whether a material possesses good or poor wear resistance is only valid when accompanied by a clear specification of the conditions under which the assessment was conducted.

Several comprehensive reviews on wear are available in the literature. For instance, Blau \cite{Blau} focuses mainly on metals, while Muhr and Robbins \cite{Muhr} and Veith \cite{Veith} have investigated rubber wear. A broader historical perspective on wear and tribology can be found in the beautiful book by Dowson \cite{Dowson}.

In this article, we focus on sliding wear but also briefly discuss fretting (a form of sliding wear) and erosion, where the surface of a solid is exposed to a beam of impacting hard particles. We begin by reviewing the development of wear theories, primarily in metals. Then, we shift focus to rubber-like materials, highlighting the most influential works by pioneers such as Schallamach, Grosch, Southern and Thomas, and Arnold and Hutchings. We also consider recent studies that attempt to include the multiscale nature of roughness on real surfaces.

\begin{figure}[tbp]
\includegraphics[width=0.47\textwidth,angle=0]{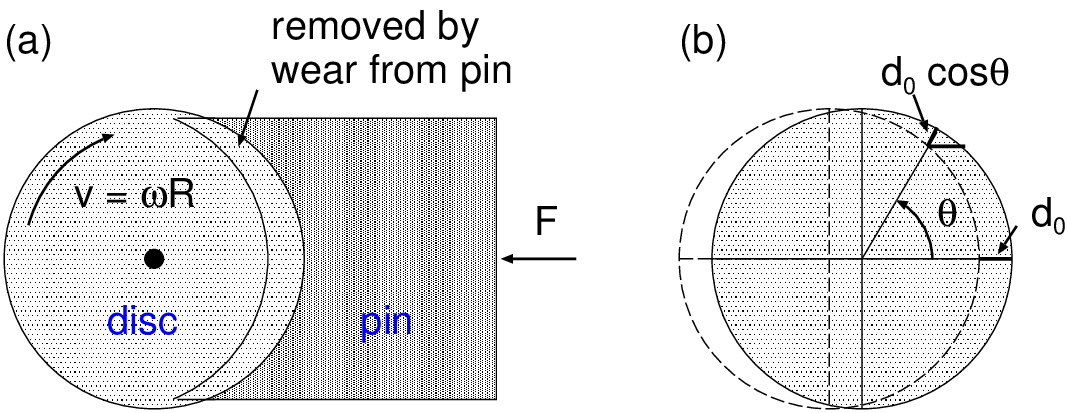}
\caption{
Wear in a brake system. A solid bar (pin) is pressed against a rotating cylinder (disc).
If the rotating disc does not wear, then the wear of the pin is equivalent to the disc 
moving toward the pin. The thickness of the removed material can be obtained 
as the difference in circumferences between two displaced discs. The pressure is assumed 
to act orthogonally to the rotating disc, and the thickness of the removed material 
in this direction is given by $d(\theta) = d_0 \cos\theta$. Hence, if the wear rate is 
proportional to the pressure, then $p \sim d(\theta) \sim \cos\theta$.
}
\label{BreakSetUp1.eps}
\end{figure}

\vskip 0.3 cm
{\bf 2 Foundations of Wear Theories}

One of the first scientific studies of wear was the work by K.T. Reye \cite{reye}. He proposed in 1860 that the wear is proportional to the frictional work. This is in general not true since adding a monolayer of a boundary lubricant will usually change the frictional work and the wear differently \cite{moli00}. In addition, the friction is usually due mainly to energy transfer into thermal movements and not due to wear. Still, the wear rate may be proportional to the (local) nominal contact pressure $p({\bf x})$, at least if frictional heating is unimportant, and since the frictional power often is proportional to the nominal contact pressure, $\mu p v$, if the friction coefficient is assumed constant then the local wear rate may be proportional to the frictional power. When the friction shear stress $\mu p$ is proportional to the pressure $p$, the area of real contact is usually proportional to $p$, so the statement that the wear rate is proportional to the frictional work is often equivalent to the statement that the wear rate is proportional to the area of real contact (see below).

Reye used this assumption to predict the pressure distribution in sliding contacts after run-in has occurred. As an example, consider the simple break in Fig. \ref{BreakSetUp1.eps}. After run-in, assuming material is removed only from the stationary object (pin), if the wear rate is proportional to the nominal contact pressure then the nominal pressure distribution must equal $p \sim {\rm cos}\theta $.

In 1953, J.F. Archard \cite{A1} proposed a wear model establishing a proportionality between the volume of wear $V$ produced during sliding, the applied load $F_0$, the area of real contact, and the sliding distance $L$:
$${V\over L} = K_0 {F_0\over \sigma_{\rm P}} \eqno(1)$$
where $\sigma_{\rm P}$ is the yield stress or penetration hardness. 
This model is now known as Archard's wear model which can also be written as
$${V\over A_0 L} = K_0 {\sigma_0\over \sigma_{\rm P}}$$
where $\sigma_0 = F_0/A_0$ is the (average) nominal contact pressure and $A_0$ the nominal contact area.

The Archard wear model assumes that the asperity contact regions have yielded plastically, so the area of real contact is taken as $A = F_0/\sigma_{\rm P}$. Hence, we may state that the wear rate $V/L$ is proportional to $A$, with the added condition that the pressure in the contact regions is determined by the penetration hardness. This condition is important for wear of metals, where it is likely that the asperity contact regions must deform plastically for cold-welded junctions to form, resulting in metal transfer or particle removal.

In a subsequent study, Archard and Hirst \cite{A2} reported comprehensive experimental on the sliding wear of metals. They concluded that under steady-state surface conditions (which correspond to kinetic equilibrium), the wear rates of materials are indeed proportional to the applied load (unless a change in load causes the surface conditions to change) and independent of the apparent area of contact. These observations are expected if the area of real contact is proportional to the load, as would be the case if all contact regions deform plastically. However, in many applications, surfaces make repeated contact with other surfaces, and after many contacts one expects only elastic deformations of the surface asperities. Furthermore, for some materials like rubber or gel, no plastic deformation is expected.

If the contact regions deform elastically rather than plastically, then there is no obvious reason for the penetration hardness to appear in the wear equation. For this case, we write
$${V\over L} = K_1 {F_0 \over E^*} \eqno(2)$$
where we have normalized the normal force with the effective 
Young’s modulus $E^*$ so that the wear coefficient $K_1$ is dimensionless.

Modern contact mechanics theories for elastic solids with randomly rough surfaces predict that the area of real contact is proportional to $1/E^*$ and to the normal force, and also independent of the nominal surface area if the surface roughness has a large enough roll-off region, i.e., the length scale $\lambda_{\rm r}$ for the onset of the roll-off region is much smaller than the linear size of the nominal contact area. The latter is almost always true for engineering objects, which are designed to have smooth surfaces on the macroscopic scale. However, the area of real contact is not independent of the nominal contact area if the roll-off length $\lambda_{\rm r}$ is on the order of, or larger than, the linear size of the nominal contact area (see Appendix A).

One ``problem'' with the Archard wear equation is that experiments show that the wear factor $K_0$ can take a huge range of values, from $10^{-7}$ to $10^{-1}$. Hence, while the proportionality of the wear rate to the normal force and the sliding distance is valid in most cases after run-in, a microscopic physical explanation for the origin of the wear coefficient $K_0$ is needed. We will present one such model later, which can explain the large range of wear coefficients observed.

The transfer of material between two surfaces during sliding has been intensively studied \cite{Green}. By the late 1940s, radioactive isotopes were becoming more readily available to researchers, making it possible to study the processes of material transfer in the wear of metals.
These techniques were applied by Rabinowicz \cite{A4} in his studies of metal wear. 
Experimental studies have shown that the wear of metals often proceeds through multiple stages. Initially, metal fragments are transferred from one surface to another. These fragments are then smeared across the surface, forming a transfer layer that is typically harder than the original material. This layer subsequently undergoes oxidation, a process accelerated by the elevated temperatures generated through frictional heating. As the oxidation progresses, the bond between the transfer layer and the underlying surface becomes weakened. Eventually, the oxidized material is removed in the form of large oxide particles, which constitute the wear debris \cite{Green}.
Wear particles can remain trapped between the surfaces and reattach several times before finally emerging
outside of the contact.
The transfer of metals in sliding contacts is extremely important in the performance of 
switches, relays, edgeboard connectors, and other electromechanical components.

Several microscopic mechanisms for the origin of wear have been presented. Nam P. Suh proposed the delamination theory of wear \cite{nam}, a sliding wear model introduced in 1973 that is based on fatigue mechanisms. This model describes how subsurface cracks initiate and grow, leading to the formation of flat wear particles that eventually separate from the surface.

\begin{figure}[tbp]
\includegraphics[width=0.25\textwidth,angle=0]{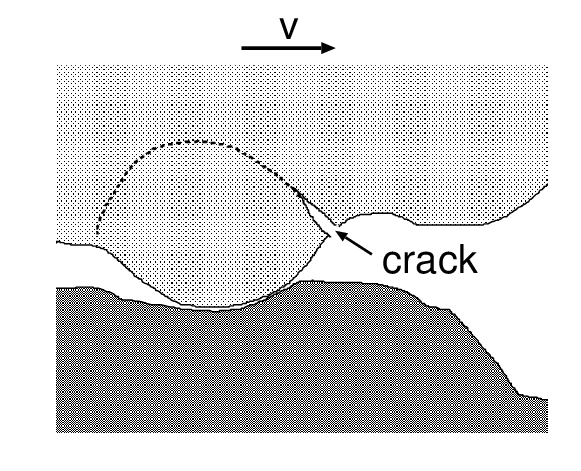}
\caption{
To remove a particle of linear size $D$, enough elastic deformation
energy (due to the stress from the countersurface)
must be stored in its vicinity to break all the bonds necessary to form the free particle.
The energy to break the bonds is $U_{\rm c} \approx \gamma D^2$, where $\gamma$ is the cohesive energy per unit surface area,
and the elastic energy scales as $U_{\rm el} \approx (\tau^2/E) D^3$, where $\tau$ is the shear stress in the asperity contact region.
Hence, removal of a particle of size $D$ is possible only if $D > \gamma E/\tau^2$. Thus, only sufficiently large asperity contact regions 
will give rise to wear particles.
}
\label{PicRab1.eps}
\end{figure}

Rabinowicz \cite{A4,A3} presented a very interesting
wear model in which wear particles are removed when
the elastic deformation energy in asperity contact regions becomes
larger than the energy needed to break the bonds between 
the particle and the surrounding material, see Fig. \ref{PicRab1.eps}.
In this model, the formation of a wear particle is due to crack propagation,
and the energetic condition used by Rabinowicz is similar to the 
fracture criterion formulated by Griffith (1921) \cite{griffith}. However, while
Griffith’s crack propagation theory describes the conditions for a crack to 
grow (involving different elastic energy release rates depending on crack size),
the Rabinowicz criterion only assumes that sufficient elastic energy is available to break the bonds
and form the wear particle, without considering the fracture process in detail.
Rabinowicz did not consider the multiscale nature of contact between solids,
but this was included in recent studies which also account for the fact that the removal of
particles may be a fatigue process, requiring many contacts 
with asperities before a particle is removed \cite{RP1,RP2,RP3}.

\begin{figure}[tbp]
\includegraphics[width=0.40\textwidth,angle=0]{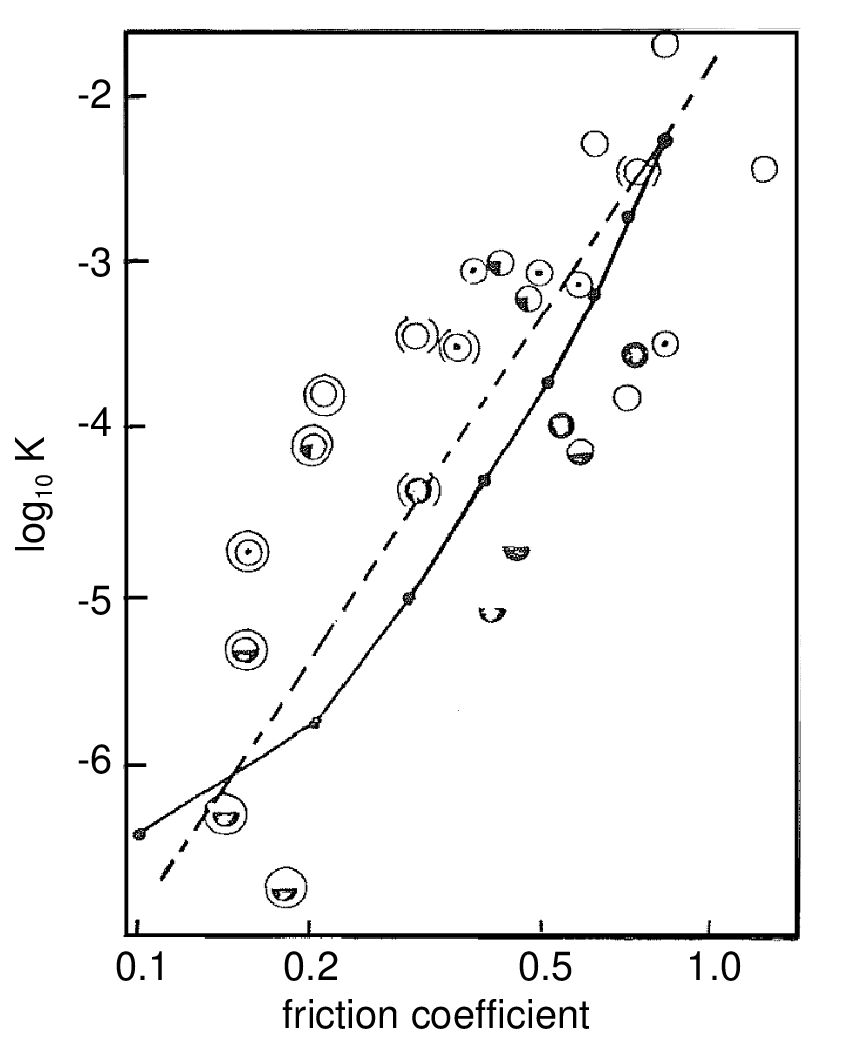}
\caption{
The wear coefficient $K_0$ as a function of the friction coefficient $\mu$ (log-log scale)
for unlubricated and lubricated metal pairs. The dashed line indicates wear proportional to
$\mu^5$. Adapted from Ref. \cite{Rab1}.
}
\label{PicRab2.eps}
\end{figure}

One important implication of the Rabinowicz theory is that if the frictional shear stress
(or the kinetic friction coefficient) is small enough, there may not be enough elastic energy to
form wear particles. Hence, a boundary lubrication film can have
a large influence on the wear rate by reducing the frictional shear stress
to such a degree that insufficient elastic energy is available to form wear particles.
In this model, a strong dependence of the wear rate on the 
sliding friction coefficient is expected, as was indeed observed 
by Rabinowicz in a large set of experiments,
see Fig. \ref{PicRab2.eps}. The Rabinowicz theory can explain the wide range of
wear rate coefficients $K_0$ observed experimentally.

\begin{figure*}[tbp]
\includegraphics[width=0.6\textwidth,angle=0]{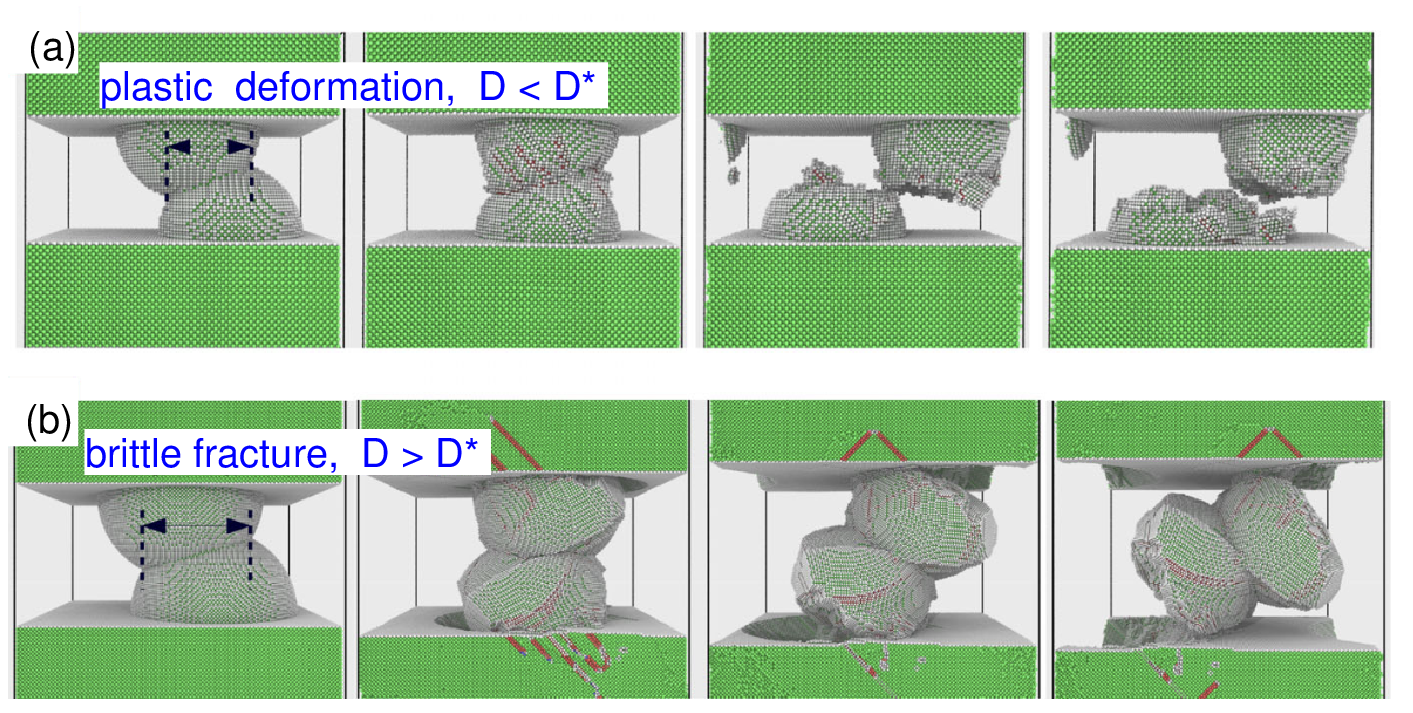}
\caption{
Numerical simulations distinguishing two wear mechanisms at the asperity level. (a) shows the plastic smoothing mechanism
in the absence of wear debris for a {\it small} junction. (b) presents the fracture-induced (crack propagation) particle formation
mechanism for a large asperity junction. 
The critical length $D^* \approx \gamma E/\tau^2$ (see Fig. \ref{PicRab1.eps}).
Adapted from Ref. \cite{Moli0}.
}
\label{PlasticFracture.eps}
\end{figure*}

In a recent series of papers, Molinari and coworkers \cite{Moli0,Moli1,Moli2,Moli3,Moli4} have studied 
wear processes using molecular dynamics computer simulations. They employed realistic 
interaction potentials and observed the transition from plastic smoothing of asperities at short 
length scales to fracture-like removal of wear particles at larger 
length scales (see Fig. \ref{PlasticFracture.eps}). The fracture removal process is in 
qualitative agreement with the predictions of the Rabinowicz wear model, but the smoothing of asperities
at shorter length scale a new important discovery.

Fretting wear, typically arises in situations such as clamped or bolted mechanical joints, stacks of components during transport, and electrical connectors within vibrating equipment. It is characterized by small amplitude oscillatory motion. Unlike reciprocating sliding wear, fretting wear often involves the trapping of debris within the contact region. Additionally, the outer areas of the contact may experience slipping, whereas the central region can remain adhered. Cattaneo \cite{Cat} and Mindlin \cite{Mind} published a famous paper providing an elastic contact analysis of the fretting problem.

There is a continuous transition from fretting wear to sliding wear as the amplitude of oscillation increases, but typically fretting wear occurs when the oscillation amplitude is below $\sim 100 \ {\rm \mu m}$. The magnitude of the wear volume, the electrical contact resistance, and the friction coefficient during fretting are all strongly dependent on the vibration amplitude.

Fretting wear may also involve corrosion and the oxidization of debris, which is known as fretting corrosion. Corrosion may be facilitated by water capillary bridges, which form spontaneously between closely spaced hydrophilic surfaces in a humid atmosphere. During the 1960s, there was increasing interest in understanding the mechanisms of erosive wear in applications such as electrical power plants, mining machinery, and jet engines. Foundational work by Finnie in 1958 and 1960 provided key insights into the erosion behavior of ductile metals subjected to hard particle impingement \cite{finnie1958, finnie1960}. The main parameters governing this erosion process include particle velocity, angle of incidence, and particle flux. For ductile metals, the erosion rate was typically observed to peak at incidence angles between approximately $15^\circ$ and $30^\circ$. It was also demonstrated that the erosion rate, defined as mass loss per unit mass of impacting particles, could be modeled as a constant multiplied by the particle velocity raised to a power $n$, with $n$ ranging from about $2$ to $2.5$ for metallic materials. Observations revealed the embedment of hard eroding particles such as alumina into the surface, the development of subsurface damage gradients, and fatigue-like dislocation structures within the affected regions of the material.

\begin{figure}[tbp]
\includegraphics[width=0.48\textwidth,angle=0]{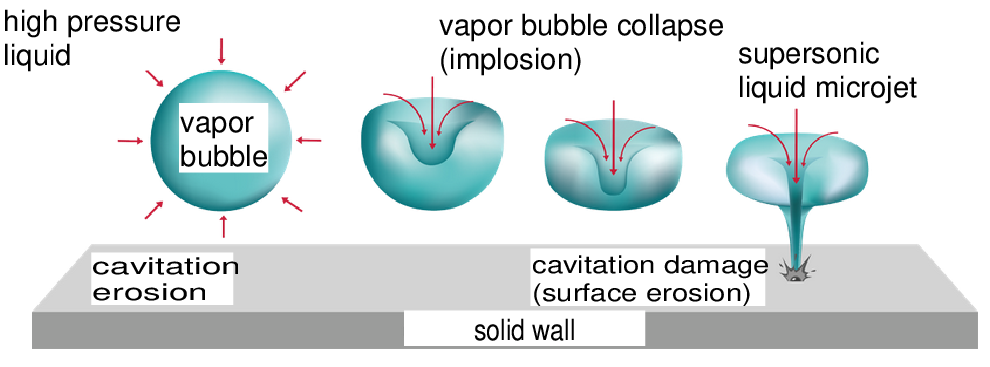}
\caption{
When a vapor bubble implodes, the bubble's center 
pulls in a jet of liquid at supersonic speed. If these 
implosions occur close to a solid wall, 
the powerful microjet will erode material from that surface.
The direction of the microjet can be due to a sound wave 
which break the symmetry of the problem.
Courtesy of Special Alloy Fabricators (SAF),
see https://www.slurryflo.com/cavitation
}
\label{CavityImplode.eps}
\end{figure}

Erosion of solid materials in liquid environments can result from the collapse of cavitation bubbles \cite{cavitation}. These bubbles form when the local liquid pressure falls below the saturated vapor pressure $p_{\rm sat}$. Under low ambient pressure, the bubbles expand, and they collapse once the surrounding liquid pressure rises above $p_{\rm sat}$. When bubble collapse occurs near a solid boundary, it can produce high-velocity liquid jets and shock waves that exert extreme pressure on the surface. These transient pressure loads are responsible for causing erosive damage to solids, as seen in systems such as liquid fuel injectors, hydropower equipment, and marine propulsion units. In contrast, the same pressure effects generated by bubble collapse are intentionally utilized in applications like shock wave lithotripsy, targeted drug delivery, and precision surface cleaning.

\begin{figure}[tbp]
\includegraphics[width=0.45\textwidth,angle=0]{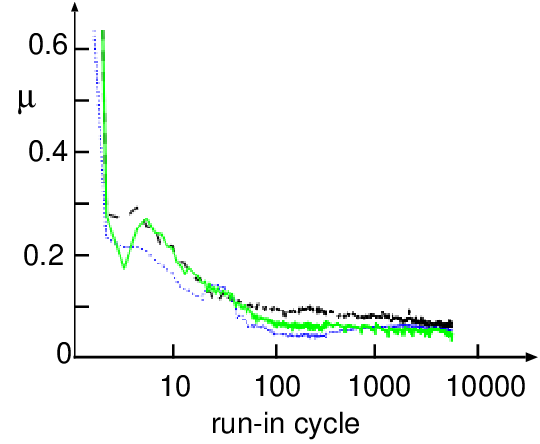}
\caption{
Friction coefficient as a function of the number of cycles for 
(metal-doped) amorphous diamond-like carbon coatings: 
TiC/a:C (dashed line), TiC/a:C-H (dotted line), and WC/a:C-H (solid line) 
sliding against steel in dry air. Adapted from \cite{Singer}.
}
\label{RunInPictureNew.eps}
\end{figure}

Many sliding contacts exhibit strongly varying friction and wear rates as a function of the sliding distance, but tend to approach a steady state after sufficiently long sliding distances (see Fig. \ref{RunInPictureNew.eps}). This phenomenon is referred to as run-in. Achieving a steady-state after an adequate run-in period is crucial in nearly all applications involving repeated use of mechanical components, such as the piston–cylinder contact in combustion engines or tire–road interactions.

Run-in typically involves the smoothing of asperities through fracture or plastic deformation. In the case of metals, it may also include material transfer between the contacting surfaces and the formation of modified surface layers.

In summary, fundamental concepts essential to understanding wear phenomena include adhesion, the true contact area, interfacial material transfer, critical angles associated with maximum abrasion and erosion, and the mechanical behavior of rough surface interactions.

\vskip 0.3cm
{\bf 3 Mechanisms of Rubber Wear}

The wear of rubber is well known to everyone from the context of passenger car tires, where darkened tracks can be found on road surfaces after braking or cornering. Tires are black because of the carbon filler, and during any form of acceleration, some slip occurs between the tire tread blocks and the road surface, resulting in the removal of small rubber wear particles with typical sizes between $1 \ {\rm \mu m}$ and $100 \ {\rm \mu m}$. This type of wear occurs even during driving at constant speed since some tread block slip also occurs due to rolling resistance and aerodynamic drag. Any torque acting on a tire will result in some tread block slip, especially near the exit of the tire-road contact patch.

In fact, just the change in the tire-road footprint as a tire rolls causes some small slip at the interface due to elastic deformation. Reducing this slip through the transition from bias-ply to radial tires was one of the key advancements in tire design. Radial tires exhibit less slip at the road-tire interface and hence much lower wear during rolling. This illustrates the earlier point that the wear resistance of a material cannot be judged without considering the specific application context.

During repeated sliding on the same surface area, the formation of a rubber transfer film will alter both the friction force and the wear rate. The influence of such films is well known in Formula 1 racing, where it is often stated that friction on a surface contaminated by a rubber film is higher than on a clean road surface. While this may be true for the sticky compounds used in F1, it is not a general rule, as shown in laboratory studies \cite{Tiwari}, where the opposite result is obtained. A decrease in wear rate on contaminated substrates has also been observed in a previous wear study by us (unpublished). Moreover, the rubber surface itself changes simultaneously with the pavement due to material transfer, thus changes in friction and wear can not be attributed solely to the formation of a rubber film on the road or substrate.

Rubber wear is believed to result from mechanical failure due to excessively high local frictional stresses, which are most likely to occur on rough tracks \cite{Muhr,S5}. The average friction level is not directly relevant to wear, one example being that friction arising from energy loss in a lubricant film does not contribute significantly to wear, as such dissipation does not generate high local strains. Another example is that a high frictional shear stress acting on a small contact area can produce the same friction force as a lower shear stress acting on a larger area, yet the wear may be significant in the former case and negligible in the latter. Therefore, what matters is the frictional shear stress acting in the contact area. Theories of abrasion therefore require detailed knowledge of local stresses, which, in combination with the strength properties of the rubber, may allow the abrasion rate to be predicted.

\begin{figure}[tbp]
\includegraphics[width=0.40\textwidth,angle=0]{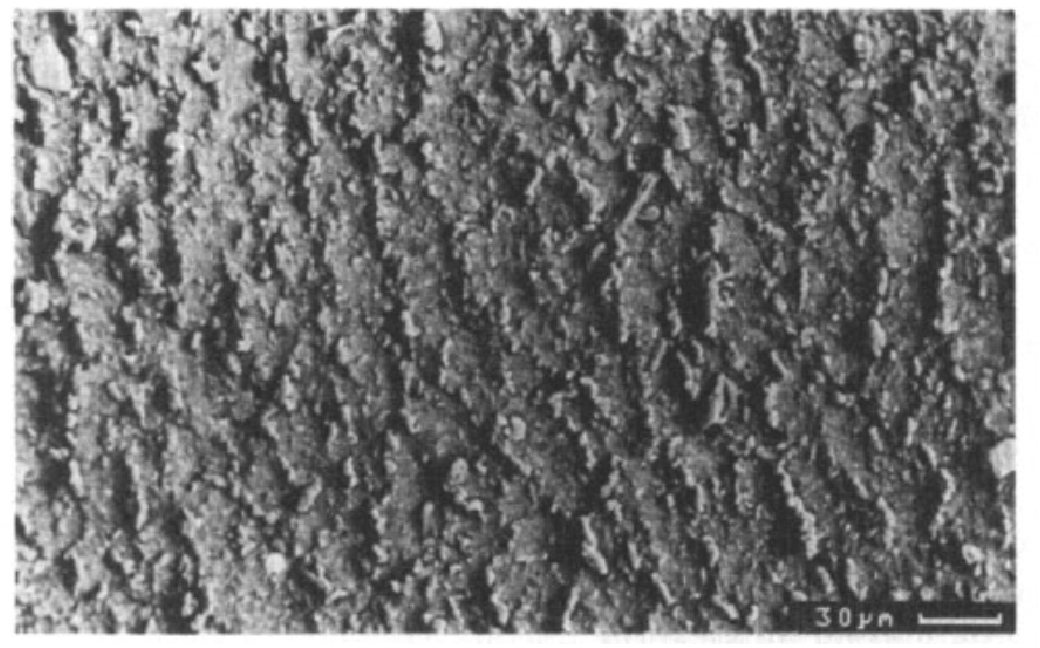}
\caption{
Scanning electron microscopy of a NR surface eroded at $30^\circ$ by $120 \ {\rm \mu m}$ silica particles at $100 \ {\rm m/s}$.
A wear pattern consisting of parallel ridges is formed on the rubber surface when a steel blade or a rough surface slides in one direction,
or when, as in the present case, a beam of hard particles impacts the surface at grazing incidence.
If the impact direction varies randomly over time, no such wear pattern is formed (intrinsic wear).
Adapted from \cite{AH0}.
}
\label{ridgesEROSION.eps}
\end{figure}

Unidirectional abrasion of rubber often results in surface patterns characterized by ridges oriented perpendicular to the sliding direction \cite{S3} (see Fig. \ref{ridgesEROSION.eps}). If the direction of sliding is continuously changed, such patterns do not form and the abrasion rate is lower. The resulting surface roughness is finer (micron-scale rather than millimeter-scale), a process referred to as ``intrinsic'' abrasion.

For sliding on rough surfaces, Grosch and Schallamach proposed the wear law \cite{S4}:
$$
{V\over L} = K_2 {\mu F_0 \over U} \eqno(3)
$$
where $U \approx \sigma_{\rm B} \epsilon_{\rm B}/2$ is the energy density at break measured at the appropriate high frequency, with $\sigma_{\rm B}$ and $\epsilon_{\rm B}$ being the stress and strain at break. This wear law suggests that the wear volume is proportional to the total frictional energy dissipated, $\mu F_0 L$. However, we do not believe this to be generally valid. A model based on the Rabinowicz picture, extended to multiasperity contact, predicts a much stronger dependence of the wear rate on the friction coefficient \cite{S5}.

Grosch and Schallamach also reported that, in laboratory studies and on road surfaces, rubber wear depends nonlinearly on the nominal contact pressure \cite{S6}:
$$
{V\over L A_0} = \left ( {p\over p_0} \right )^n
$$
where $p_0$ and $n$ are empirical constants, with $n > 1$. However, this contradicts detailed experiments we have performed on concrete surfaces at low sliding speeds. In our tests (conducted at $v \leq 1 \ {\rm cm/s}$ to minimize frictional heating), we observed a linear dependence of $V/L$ on the nominal pressure up to $p = 0.43 \ {\rm MPa}$, which is typical in the tire-road contact patch. This linear relation is also predicted by our wear theory \cite{RP1} for $p < 0.5 \ {\rm MPa}$. We believe the nonlinear behavior observed by Grosch may be due to frictional heating rather than an intrinsic property of the rubber.

In the literature, three very different types of experiments have been used to study rubber wear:

\begin{itemize}
    \item[(a)] Sliding a sharp needle or steel blade orthogonally across a flat rubber surface;
    \item[(b)] Erosion, where the rubber surface is bombarded by hard particles;
    \item[(c)] ``Normal'' wear experiments, in which a rectangular rubber block is slid on a rough surface (e.g., concrete or sandblasted steel), or a rubber wheel rotates at speed $v_{\rm R}$ on a larger rotating disk at speed $v$, corresponding to slip $s = (v - v_{\rm R})/v$. In the latter setup, the substrate often consists of a corundum disk.
\end{itemize}

We will briefly consider the wear mechanisms involved in each of these experimental approaches.

\begin{figure}[tbp]
\includegraphics[width=0.45\textwidth,angle=0]{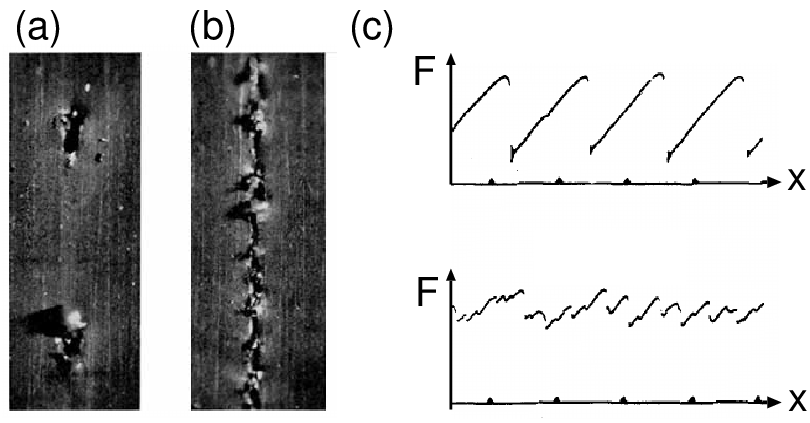}
\caption{
A gramophone needle sliding on NR without (a) and with (b) carbon filler. 
(c) In case (a), stick-slip motion is observed (top), while in case (b), the motion is more continuous (bottom). 
The normal load is $0.5 \ {\rm N}$ and the sliding speed is $3.4 \ {\rm cm/s}$. 
Adapted from Ref. \cite{S1}.
}
\label{NeedlePictures.eps}
\end{figure}

\begin{figure}[tbp]
\includegraphics[width=0.45\textwidth,angle=0]{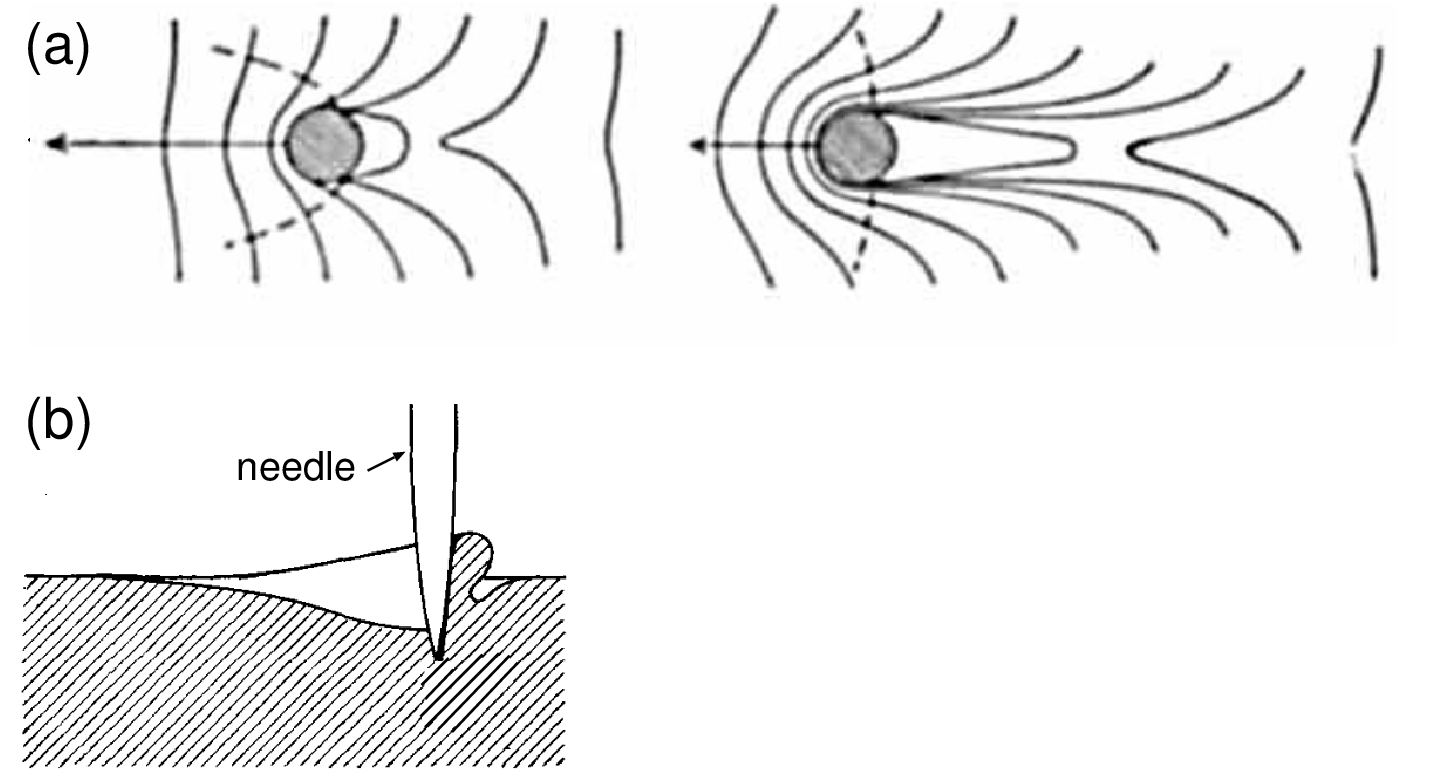}
\caption{
(a) Surface displacement field on a rubber surface as a needle is displaced to the left. 
(b) Vertical cross-section through the needle and lip corresponding to the deformation stage shown in (a). 
Adapted from Ref. \cite{S1}.
}
\label{SCHpicAB.eps}
\end{figure}

\vskip 0.2cm
{\bf Sliding of needle}

When rubber comes into contact with a surface featuring sharp roughness, very high stresses occur in the contact regions. To simulate this, A. Schallamach \cite{S1} studied the friction and wear resulting from sliding a gramophone needle on a natural rubber (NR) surface. This represents one of the earliest experimental studies of rubber wear. The surface damage observed in this experiment depends strongly on the nature of the rubber. In particular, the traces left on unfilled NR are discontinuous and exhibit stick-slip characteristics, whereas traces on a NR tire tread compound with carbon filler appear more continuous (see Fig.~\ref{NeedlePictures.eps}).

When a needle is pressed into rubber, it creates a funnel-shaped depression. If the applied force is small, this deformation may be primarily viscoelastic and not involve bond breaking. Schallamach showed that as the needle moves, its leading edge contacts the sidewall of the funnel, causing the rubber to bulge in front of the needle [see Fig. \ref{SCHpicAB.eps}(b)]. The elastic force acting on the needle has a vertical component directed upward, causing the needle to lift the rubber adhering to it. The bulge forms a lip. As the needle moves over the lip, the rubber retracts, corresponding to the tip jumping forward on the surface, producing the stick-slip motion and force fluctuations shown in Fig. \ref{NeedlePictures.eps}(c) (top).

In Ref. \cite{Julia}, an engineering stylus was slid across a flat silicone rubber surface. Although the force was not measured, the vertical tip position displayed stick-slip oscillations similar to those in Fig. \ref{NeedlePictures.eps}(c). In that case, no bond breaking may have occurred, but if the normal load is high enough, sufficient elastic energy may accumulate at the needle tip to cause crack initiation and irreversible surface damage, as observed in Figs. \ref{NeedlePictures.eps}(a) and (b).

Rubber filled with carbon is stiffer than unfilled rubber, resulting in higher stresses and greater elastic energy storage, approximately $E\epsilon^2/2$, for the same strain field. As a result, fracture may occur before lift-off over the lip, leading to a more continuous wear track, as seen in Fig.~\ref{NeedlePictures.eps}(b).

\begin{figure}[tbp]
\includegraphics[width=0.48\textwidth,angle=0]{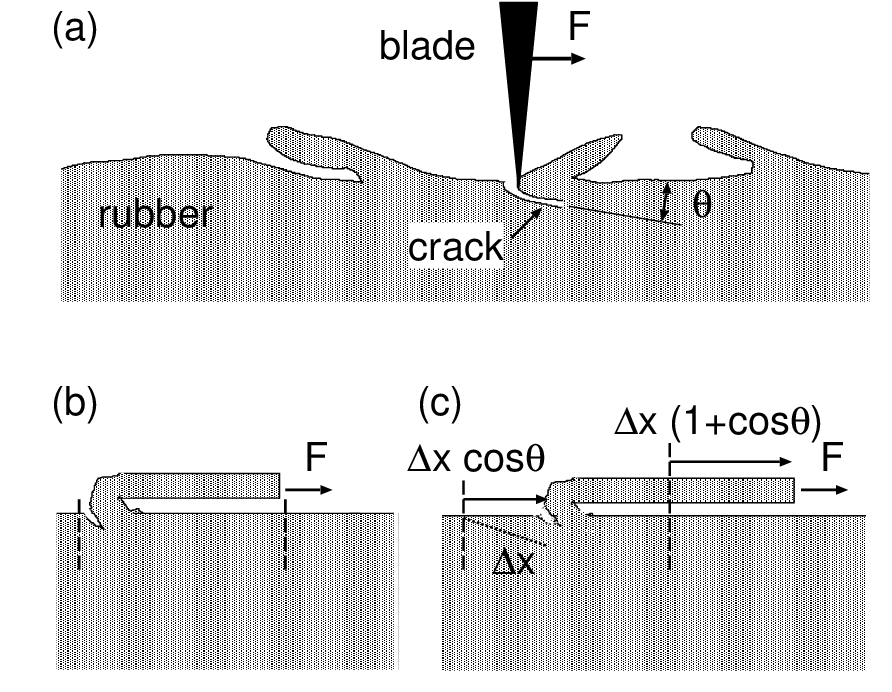}
\caption{
(a) Model for crack growth: the blade moves along the surface, and a tangential force $F$ acts on the rubber tongue. The crack propagates at an angle $\theta$ to the surface. 
(b) and (c) When the crack propagates a distance $s$, the external force performs work $F s (1 + \cos\theta)$.
}
\label{tangue1.eps}
\end{figure}

\vskip 0.2cm
{\bf Sliding of blade}

Southern and Thomas \cite{S7} presented a fundamental study on rubber wear in which, instead of a needle, they used a razor blade. They observed that scraping a rubber surface with a razor blade, after a run-in period, resulted in a wear pattern consisting of a series of ridges oriented transversely to the sliding direction (see Fig.~\ref{ridgesEROSION.eps}). Run-in typically involved multiple contacts between the blade and the rubber, and rubber films of thickness equal to the pattern height had to be removed by abrasion before the steady state was reached.

Southern and Thomas developed a theory relating the abrasion rate to crack growth at the base of the ``tongues'' of the ridges (see Fig.~\ref{tangue1.eps}). In this theory, the abrasion rate is linked to the crack growth properties of the rubber, the angle of crack growth $\theta$, and the tangential friction force $F$ acting on the blade. The theory successfully describes the behavior of non-crystallizing rubbers. However, for natural rubber (NR), the observed abrasion rate was higher than expected from its excellent crack growth resistance. The superior crack resistance of NR is attributed to its ability to strain crystallize. The apparent ineffectiveness of strain crystallization under abrasion conditions may be due to the short contact time between the rubber and the moving blade. It was also found that lubricants significantly reduce the abrasion rate, indicating that frictional stress, rather than cutting, is primarily responsible for the wear process \cite{S7}.

The energy required to propagate a crack a distance $\Delta x$ is given by the fracture energy $G$ as $\Delta U = G w \Delta x$, where $w$ is the width of the crack. When the crack has propagated a distance $\Delta x$, the blade has displaced by $\Delta x(1+\cos\theta)$ [see Fig.~\ref{tangue1.eps}(b,c)], so the work done by the force is $F \Delta x (1+\cos\theta)$. Equating this work with the energy needed to create the crack surface area $w \Delta x$ gives
$$G = \frac{F}{w}(1+\cos\theta). \eqno(4)$$

When a rubber block with a crack is subjected to an oscillating driving force, the crack grows a distance $\Delta x$ during each oscillation period. Experiments have shown that for many materials, $\Delta x$ is related to the maximum $G$ value attained during each cycle as:
$$\Delta x = B G^n \eqno(5)$$
This $\Delta x(G)$ law is often referred to as the Paris law, although it was first proposed for rubber by Thomas. The exponent $n$ typically ranges between 2 and 4. The crack propagation law (5) depends on the frequency of
the oscillating driving stress but this is seldom studied and 
usually not taken into account in an accurate way in applications.

As the blade moves a distance $L$, the volume of rubber removed is $\Delta V = w L \Delta x \sin\theta$, so the wear rate becomes
$${\Delta V \over L} = w \, \Delta x \, \sin\theta = w \, \sin\theta \, B \, \left[\frac{F}{w}(1+\cos\theta)\right]^n. \eqno(6)$$
This relation implies that ${\rm log}(\Delta V/L)$ should scale linearly with $n \, {\rm log} F$, which is consistent with experimental observations. Analysis of experimental data yields crack growth angles $\theta$ in the range of $1^\circ$ to $15^\circ$.

It is worth noting that the radius of curvature of the blade does not appear in (6), which is consistent with the observation that the wear rate is the same when using a new razor blade as when using one that has been employed for several thousand revolutions \cite{sharpness}. This supports the conclusion that cutting plays a negligible role in the abrasion mechanism under these conditions.

\vskip 0.2cm
{\bf Erosive wear from the impact of hard particles}

It is well established that resilient elastomers can provide excellent resistance to erosive wear, and materials of this type are widely used in applications such as ore-treatment plants and pipe linings. The erosive wear of rubber was studied in two fundamental papers by Arnoldt and Hutchings \cite{AH0,AH1}. They developed a model to predict erosive wear caused by the impact of small hard particles. 

Previous work had shown that the erosive wear mechanism of elastomers under glancing impact is similar in nature to abrasive wear caused by a sharp blade. The model developed by Arnoldt and Hutchings was based on the sliding abrasion model of Southern and Thomas. They combined the crack growth formulation from Southern and Thomas with a model of particle impact, assuming a force-displacement relationship for the particle-rubber interaction
corresponding to a flat-ended punch indenting a semi-infinite elastic solid.

\begin{figure}[tbp]
\includegraphics[width=0.40\textwidth,angle=0]{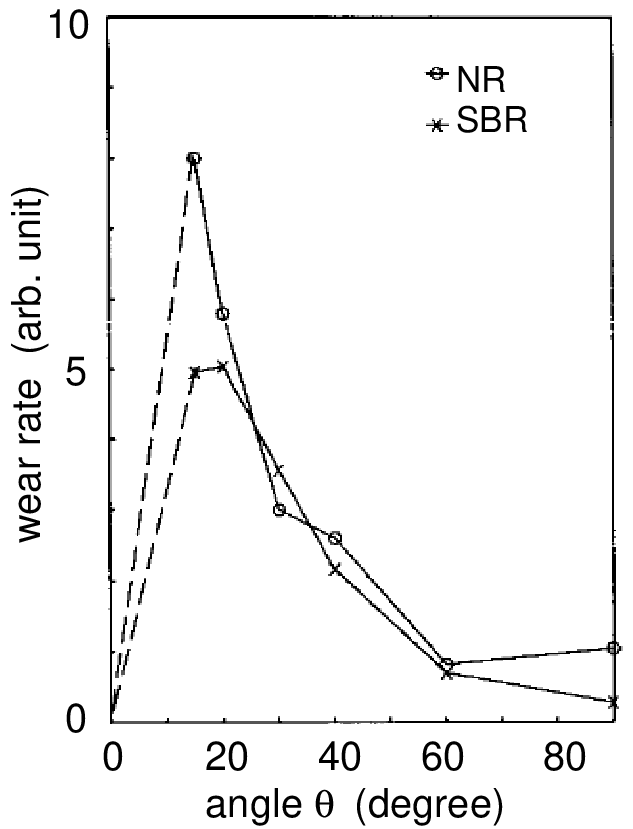}
\caption{
The variation of erosion rate with impact angle for NR and SBR eroded by
$280 \ {\rm \mu m}$ silica particles at $95 \ {\rm m/s}$ impact velocity. 
Adapted from \cite{AH0}.
}
\label{WearRateErosion.eps}
\end{figure}

Depending on the angle of incidence of the hard particles, two distinct wear mechanisms operate: one dominates at glancing angles of impact, and the other becomes significant under conditions of normal impact. In both cases, material is removed from the surface through fatigue crack propagation.

\vskip 0.1cm
{\bf Glancing incidence}—Here we first present the basic concepts for the glancing impact, a case in which the mechanism of material removal is similar to that observed in sliding abrasion by a blade.

\begin{figure}[tbp]
\includegraphics[width=0.48\textwidth,angle=0]{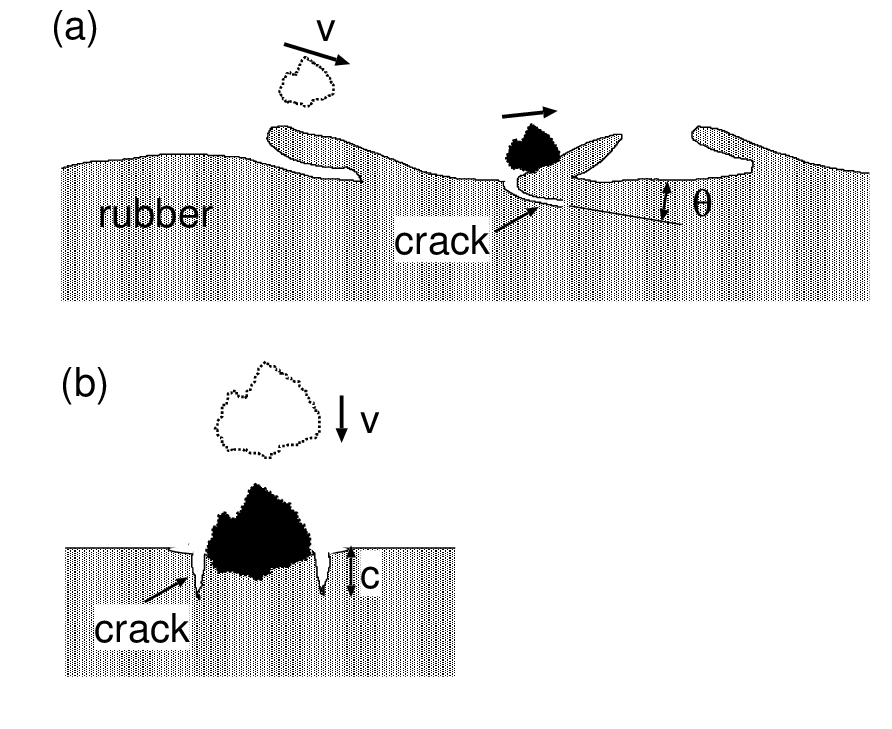}
\caption{
Erosive wear caused by hard particles at (a) glancing incidence and (b) at normal incidence.
}
\label{tangueerosion.eps}
\end{figure}

Arnoldt and Hutchings \cite{AH0} observed that, during erosion under glancing impact conditions, a series of ridges oriented transverse to the impact direction is formed during the initial run-in phase. This pattern is similar to that produced by scraping with a razor blade and is illustrated in Fig. \ref{ridgesEROSION.eps}. As the particles impact and slide along the surface, the ridges get deformed, leading to the propagation of fatigue cracks from the base of each ridge, as shown in Fig. \ref{tangueerosion.eps}(a). Similar to blade-induced abrasion, the propagation of these fatigue cracks is considered to be the rate-determining step in the erosive wear process.

They adapted the model developed by Southern and Thomas \cite{S5} to describe blade abrasion. In order to extend this model to the case of erosion, it is necessary to estimate the frictional force exerted by each impacting particle on the surface and the distance over which the particle slides during the contact time. While these parameters can be experimentally controlled or measured in blade abrasion, they are governed by the dynamics of impact in erosive conditions.

\begin{figure}[tbp]
\includegraphics[width=0.40\textwidth,angle=0]{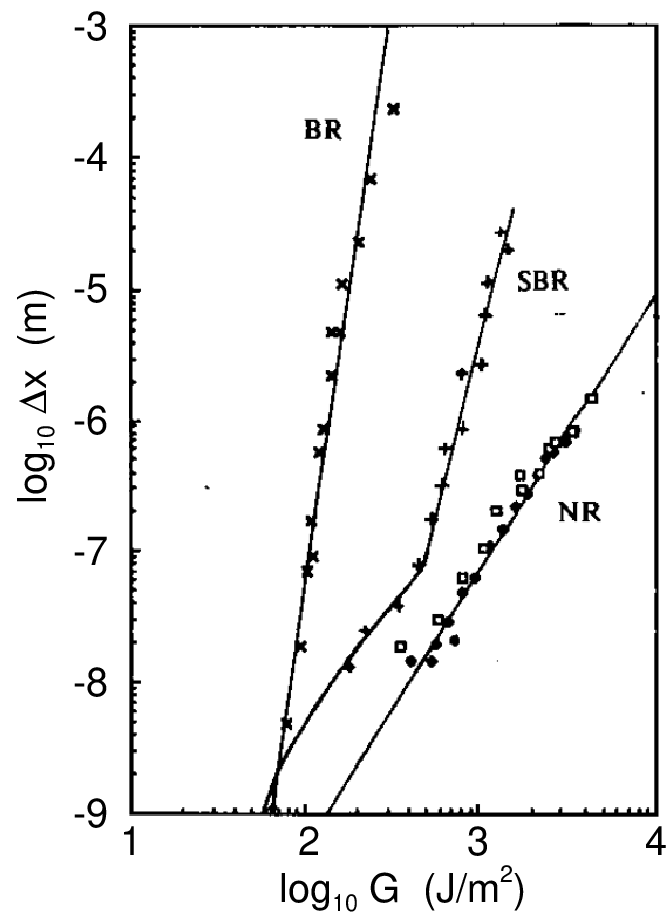}
\caption{
The displacement $\Delta x$ of the crack tip during one cycle of oscillatory stress as a function of the maximal fracture or tearing energy $G$. Adapted from Ref. \cite{AH0}.
}
\label{logGlogDxRelation.eps}
\end{figure}

Arnoldt and Hutchings used Eq. (6) to calculate the volume of eroded rubber per impacting particle:
$$
\Delta V = w L \, {\rm sin}\theta \, B \left[ \frac{F}{w}(1 + {\rm cos}\theta) \right]^n.
$$
To apply this equation, one needs to determine the force $F$ a particle exerts on a ridge, as well as the width $w$ and the length $L$ over which the particle slides on the ridge. They estimated $F$ and $L$ by solving Newton's equations for an impacting particle, assuming that the normal force can be derived from a rigid flat punch model for the particle-solid interaction and that the tangential (frictional) force is given by $F = \mu F_0$, where $\mu$ is the friction coefficient and $F_0(t)$ the normal force. For the contact width, they assumed that the normal force is high enough that $w \approx 2R$, where $R$ is the particle radius. Using this simple theory along with the measured $\Delta x(G)$ relation for several rubber types (see Fig. \ref{logGlogDxRelation.eps}), they were able to estimate the dependence of erosion rate on the angle of incidence, impact velocity, particle size, and rubber elastic modulus. These dependencies are all in semi-quantitative agreement with experimental data. The predicted absolute values of the erosion rate agreed with experiments within a factor of $\sim 10$.

\begin{figure}[tbp]
\includegraphics[width=0.3\textwidth,angle=0]{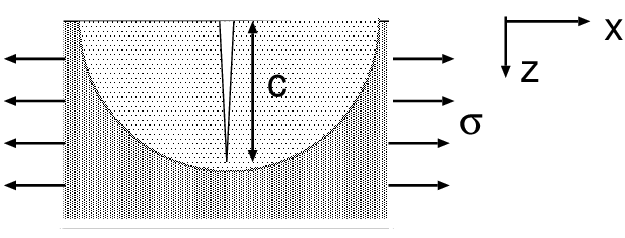}
\caption{
A crack extending a distance $c$ into an elastic block exposed to an external tensile stress $\sigma$ will reduce the elastic energy to near zero in a volume $\sim c^2 s$ (light dotted area), where $s \gg c$ is the crack length in the $y$-direction parallel to the surface.
}
\label{CrackReduceElast.eps}
\end{figure}

\vskip 0.1cm
{\bf Normal incidence}—At high impact angles, tensile stresses are generated in the surface due to particle impacts, causing fine fatigue cracks to progressively grow into the material [see Fig. \ref{tangueerosion.eps}(b)]. Material loss occurs where these cracks intersect.

Assuming again that the normal stress $\sigma_z$ acting on the rubber surface from the impact of a particle is given by the flat-ended punch model, Arnoldt and Hutchings \cite{AH1} estimated the radial tensile stress as
$$\sigma_r \approx \mu \sigma_z\eqno(7)$$
for $r \approx R$, and used this to estimate the elastic energy release rate that drives the crack growth:
$$G \approx c {\sigma_r^2 \over E} \eqno(8)$$
where $c$ is the instantaneous crack length. (Here, we neglect factors of order unity included in Ref. \cite{AH1}.) This expression is based on the assumption that a crack extending into the solid a distance $c < R$ will reduce the elastic energy in a volume of approximately $c^2 2\pi R$ (see Fig. \ref{CrackReduceElast.eps}). If the elastic energy density before the crack was formed is $\sigma_r^2/2E$, the reduction in elastic energy is $U_{\rm el} \approx \pi c^2 R \sigma_r^2 / E$. The energy required to create the fracture surface area $2\pi R c$ is $U_{\rm s} = 2\pi R c G$. At equilibrium, $dU_{\rm el}/dc = dU_{\rm s}/dc$, leading to the condition $G \approx c \sigma_r^2 / E$.

Combining Eqs. (5), (7), and (8), we obtain
$$\Delta x = B \left ( \frac{c (\mu \sigma_z)^2}{E} \right )^n \eqno(9)$$

Each particle impact will induce the growth of a cylindrical crack [see Fig. \ref{tangueerosion.eps}(b)] of radius $\approx R$ by a distance $\Delta x$ into the surface. The amount of material removed depends on the density of cracks present in the surface layer. While the crack density and average crack length in steady state are not predicted by theory, they were estimated from experimental observations. The average crack length was found to be $c \approx 0.67 R$, and the total area of crack surface per unit volume in the cracked surface layer was estimated to be approximately $17/R$.

Using these estimates, Arnoldt and Hutchings were able to semi-quantitatively reproduce the observed dependence of the wear rate on the impact velocity, particle size, and the elastic modulus of the rubber. The predicted absolute wear rate agreed with the experimental results to within a factor of $10$.

\vskip 0.2cm
{\bf Sliding of rubber block on hard randomly rough substrate}

We now consider the sliding of a rubber block on a hard randomly rough substrate surface, e.g., a tire tread block on a concrete or asphalt road surface. For this case, two different wear processes have been observed: the formation of a “dry” powder of rubber particles, or the formation of a smear film. In the former case, the wear particles can be removed from the sliding track using a soft brush, while in the latter case the film adheres strongly to the substrate. We first consider the case of smear film formation.

\vskip 0.1cm
{\bf Smear film}—In contrast to abrasion occurring on sharp tracks, the abrasion rate on blunt tracks, such as concrete or worn grinding wheels, is highly sensitive to the presence of antioxidants and to the surrounding atmosphere (e.g., oxygen or nitrogen). These factors influence the abrasion process in a way that is comparable to their effects on fatigue behavior \cite{corr}.

It is well known that under certain conditions the rubber surface becomes tacky during abrasion experiments, drum testing of tires, and sometimes for tires on the road. It has been suggested that thermomechanical degradation of the polymer into a material of lower molecular weight could be responsible. That is, degradation may result from stress-aided, thermally activated bond breaking \cite{smearing}.

It is well known that, under specific conditions, rubber surfaces can become tacky during abrasion experiments, tire drum testing, and occasionally during real road tire usage. One proposed explanation is that thermomechanical degradation of the polymer leads to the formation of lower molecular weight species. In this context, the degradation is believed to result from stress-augmented thermal activation bond breaking \cite{smearing}.

At high sliding speeds during the skidding of a vehicle with locked wheels, frictional heating certainly causes degradation. However, the phenomenon of “smearing” discussed in this subsection is associated with conditions of mild abrasion, such as on smooth surfaces, and can occur even at low sliding speeds.

Experiments show that smearing can be prevented for tire tread compounds by performing abrasion tests in a nitrogen atmosphere. The most plausible mechanism of smearing appears to be the oxidative consummation of chain scissions produced by stress-aided, thermally activated processes \cite{S9}. When smearing occurs, the rate of abrasion is reduced, presumably because the smear acts as a lubricant, thereby reducing the frictional shear stress. When abrasion is low in air due to smearing, it may become greater in nitrogen because the smearing mechanism is suppressed.

\begin{figure}
\includegraphics[width=0.47\textwidth,angle=0.0]{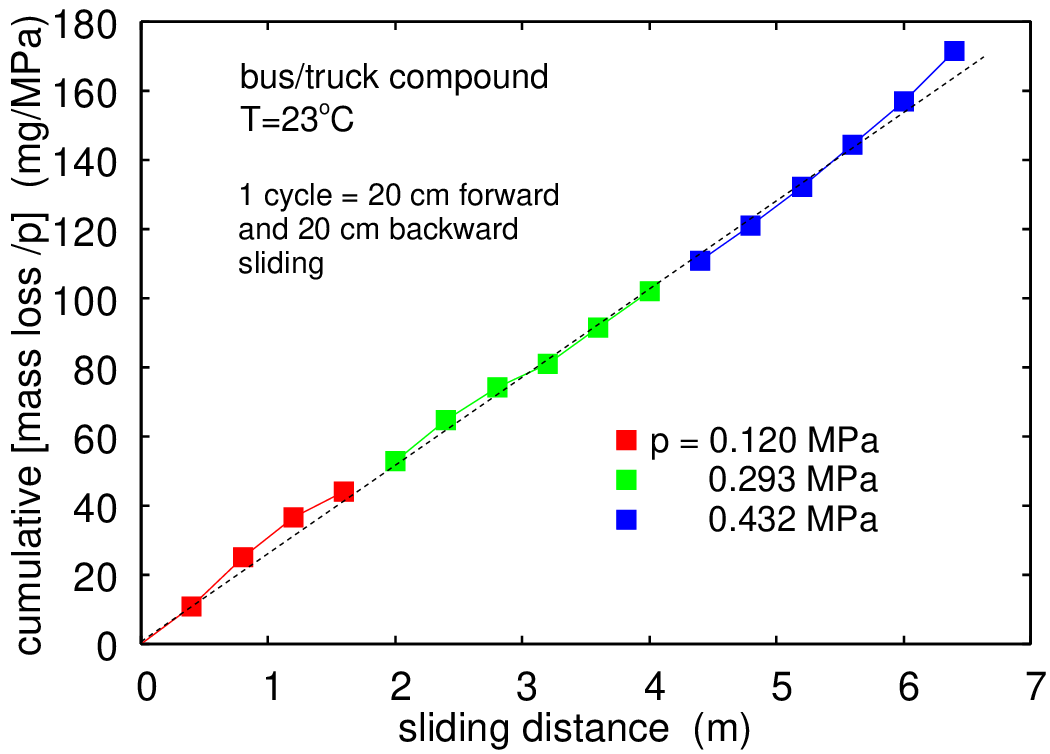}
\caption{
Cumulative wear mass $m$ over pressure $p$ in ${\rm mg/MPa}$ as a function of the sliding distance. Each data point corresponds to one sliding cycle ($20 \ {\rm cm}$ forward and $20 \ {\rm cm}$ backward motion) on a fresh concrete surface (new concrete block for each cycle). Red, green, and blue symbols represent nominal contact pressures $p = 0.120$, $0.293$, and $0.432 \ {\rm MPa}$, respectively. Sliding speed is $v = 1 \ {\rm mm/s}$. Adapted from \cite{RP1}.
}
\label{1distance.2cumulativeWear.all.pressures.eps}
\end{figure}

\vskip 0.1cm
{\bf Dry powdery wear}—We now consider the more typical case of wear involving the production of dry powdery rubber particles. The wear of rubber blocks (or rubber wheels) sliding or slipping on hard countersurfaces of various materials has been investigated in numerous studies. Here, we report on recent measurements and the theoretical framework developed in Refs. \cite{RP1,RP2,RP3}.

We studied the wear rate (mass loss per unit sliding distance) of tire tread rubber compounds sliding on concrete paver surfaces under both dry and wet conditions. Fig. \ref{1distance.2cumulativeWear.all.pressures.eps} shows that the wear rate divided by the nominal contact pressure at low sliding speed ($v = 1 \ {\rm mm/s}$) is proportional to the normal force over the nominal pressure range $\sigma_0 = 0.12$, 0.29, and 0.43 MPa. This result contrasts with the findings of Grosch, whose experiments were conducted at much higher sliding speeds (and on sharper surfaces), where frictional heating becomes significant.

\begin{figure}
\includegraphics[width=0.45\textwidth,angle=0.0]{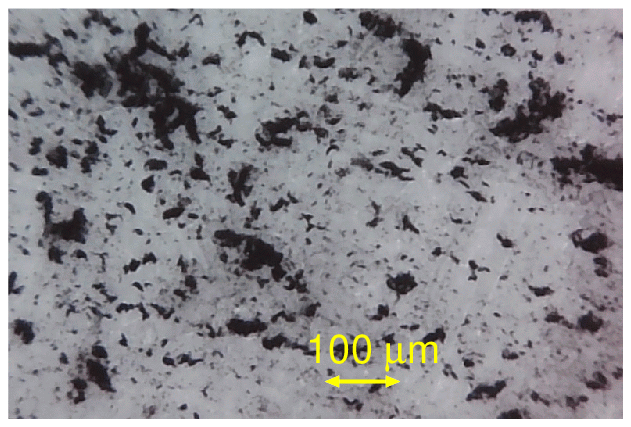}
\caption{
Rubber wear particles collected using adhesive tape after sliding a rubber block on a concrete surface. The largest particles may be agglomerates of smaller ones. Adapted from \cite{RP1}.
}
\label{RubberParticle2.ps}
\end{figure}

We also found that the wear rate is independent of the sliding velocity in the range $v = 10 \ {\rm \mu m/s}$ to $1 \ {\rm cm/s}$. A detailed study of the size distribution of the rubber wear particles was not performed, as particle agglomeration at the sliding interface complicates the analysis. However, wear particles collected using adhesive tape showed a size range of approximately $1$–$100 \ {\rm \mu m}$ (see Fig. \ref{RubberParticle2.ps}).

To explain the experimental results, we developed a wear model that incorporates the multiscale nature of real surface roughness and assumes a fatigue-based wear mechanism. The model in Ref. \cite{RP1} assumes elastic contact; in Ref. \cite{RP2}, the theory is extended to include plasticity relevant for most non-rubber materials. The model builds upon Rabinowicz’s concept, where wear particles form when the stored elastic energy in an asperity contact region exceeds the energy required to break the bonds needed to detach the particle.

\begin{figure}
\includegraphics[width=0.30\textwidth,angle=0.0]{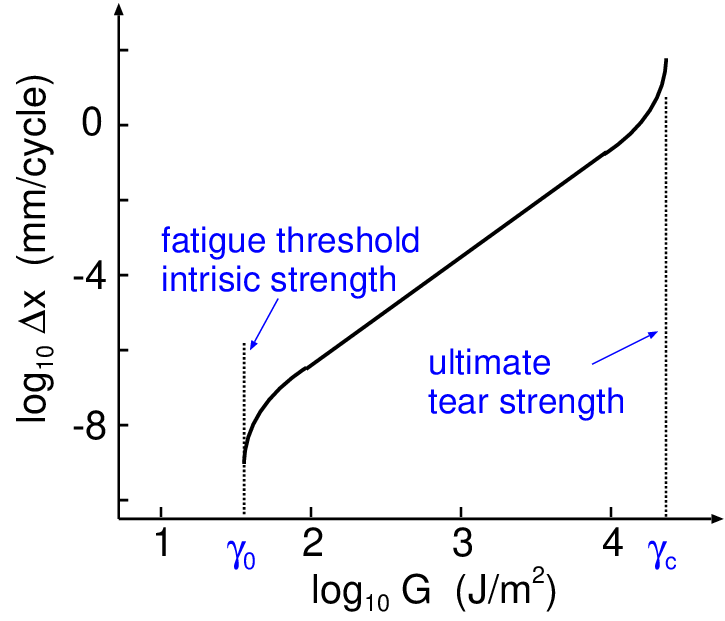}
\caption{\label{TearStrength.eps}
Crack growth length $\Delta x$ as a function of the tearing energy $\gamma$ (log-log scale). The crack is subjected to oscillating strain at $\sim 10 \ {\rm Hz}$, and the energy input per cycle is $\gamma \Delta A$, where $\Delta A = w \Delta x$ is the increase in crack surface area. Adapted from \cite{RP1}.
}
\end{figure}

However, unlike Rabinowicz’s assumption, cracks usually propagate incrementally. The crack growth per stress cycle is $\Delta x$, which depends on the tearing energy amplitude $G$ (see Figs. \ref{logGlogDxRelation.eps}, \ref{TearStrength.eps}). Thus, multiple asperity contacts are generally needed to remove a particle, except when 
the energy release rate reaches the ultimate tear strength $G_{\rm c}$, allowing single-contact detachment (see below). This is explicitly included in the model in Ref. \cite{RP1}.

The theory evaluates the interface at different magnifications. At magnification $\zeta$, the smallest observable asperities have linear size $\sim \lambda_0/\zeta$, where $\lambda_0$ is a reference length such as the nominal contact region size. The Persson contact mechanics theory is used to calculate the surface stress distribution $P(\sigma,\zeta)$, which is then used to evaluate the distribution of (temporarily) 
stored elastic energies. If the stored energy exceeds the bond-breaking energy, crack propagation may occur. The predicted wear rate is

$$
{V\over A_0 L} = {1\over 2 \ln 2} \int_{q_0}^{q_1} dq  
\int_{\sigma_{\rm c} (\zeta)}^\infty d\sigma \ 
{P(\sigma,\zeta) \over q+\pi/\Delta x(\sigma,\zeta)} \eqno(20)
$$

where $\zeta = q/q_0$ with $q_0 = 2\pi /\lambda_0$. Here $\sigma_{\rm c}(\zeta)$ is the minimum stress needed to detach a particle of size $\sim \lambda_0/\zeta$. The model also predicts the particle size distribution.

\begin{figure}
\includegraphics[width=0.47\textwidth,angle=0.0]{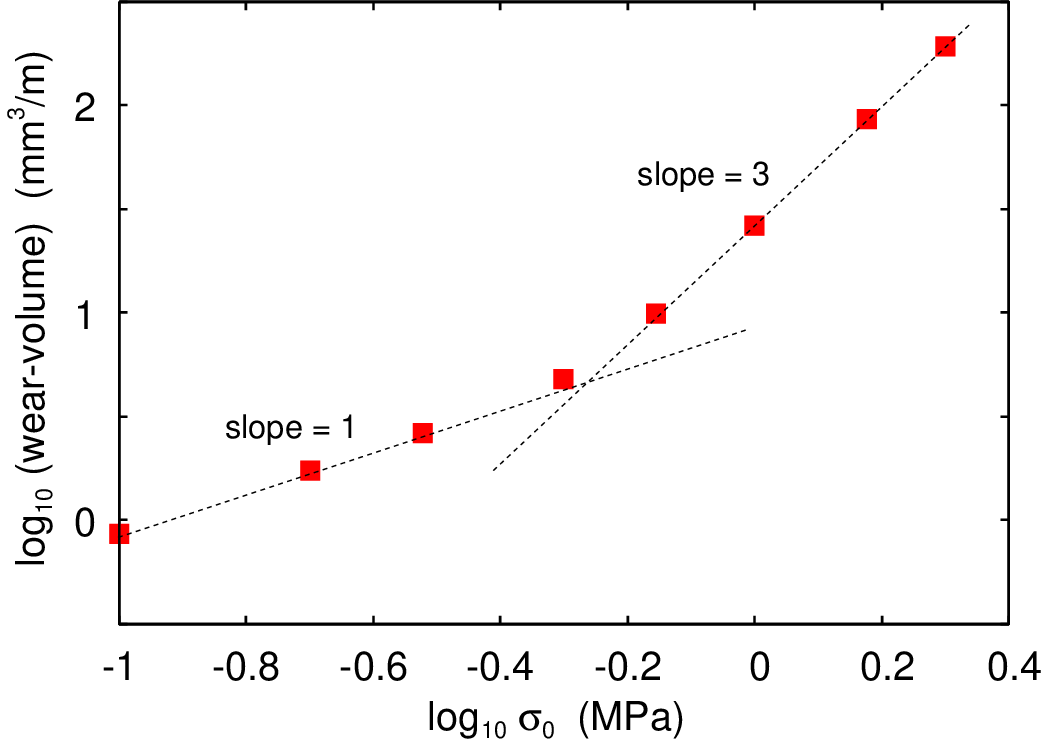}
\caption{\label{1logSigma0.2logWearRate.eps}
Calculated wear rate as a function of nominal contact pressure (log-log scale) for a rubber block sliding on a concrete surface. Adapted from \cite{RP1}.
}
\end{figure}

\begin{figure}
\includegraphics[width=0.47\textwidth,angle=-90.0]{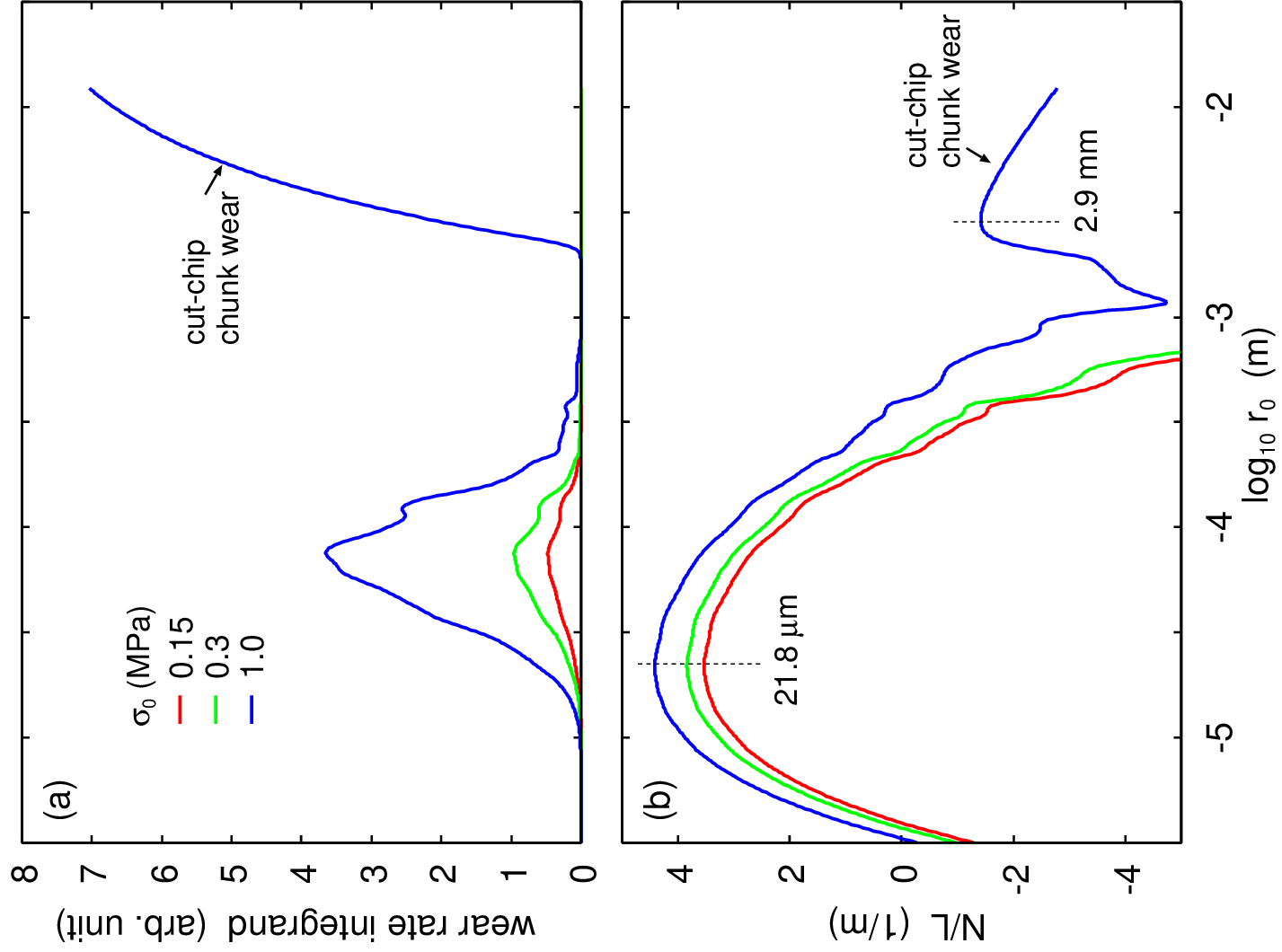}
\caption{\label{1logr0.2WearRateIntegrand.eps}
(a) Integrand of the $\xi$-integral in the wear volume expression (Eq. 21) as a function of the logarithm of wear particle radius $r_0$. (b) Particle distribution assuming a nominal contact area $A_0 = 20 \ {\rm cm}^2$. At $r_0 \approx 2.9 \ {\rm mm}$, a particle is removed every $26 \ {\rm m}$ of sliding; for $r_0 \approx 1 \ {\rm cm}$, this extends to nearly $1 \ {\rm km}$. Results shown for concrete surface under $\sigma_0 = 0.15$, 0.30, and $1.0 \ {\rm MPa}$. Adapted from \cite{RP1}.
}
\end{figure}

\begin{figure}
\includegraphics[width=0.4\textwidth,angle=0.0]{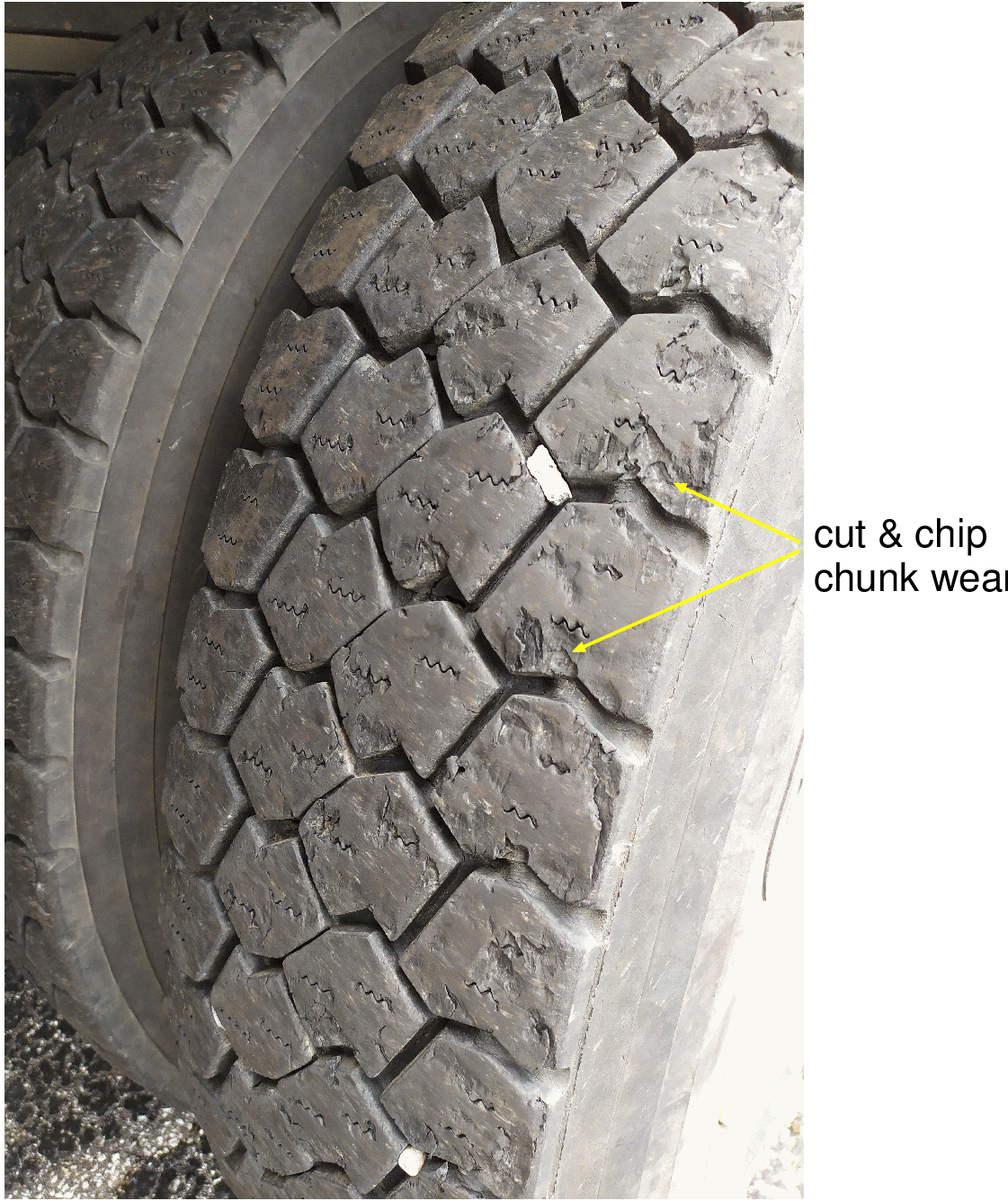}
\caption{\label{cutANDchip.eps}
Cut-and-chip wear on a truck tire due to macroscopic ($\sim {\rm cm}$) rubber fragment removal when driving over very rough surfaces such as gravel or embedded roots. Adapted from \cite{RP1}.
}
\end{figure}

The model predicts that for tread rubber sliding on concrete, the wear rate increases linearly with nominal contact pressure up to $\sigma_0 \approx 0.5 \ {\rm MPa}$. At higher pressures, the wear rate increases faster, approximately as $\sigma_0^3$ (see Fig. \ref{1logSigma0.2logWearRate.eps}). This rapid increase is attributed to the detachment of large rubber fragments, typically centimeter-sized, which is also seen in the particle size distribution (Fig. \ref{1logr0.2WearRateIntegrand.eps}). While the primary peak remains at $\sim 20 \ {\rm \mu m}$, a secondary peak at $\sim 3 \ {\rm mm}$ appears at $\sigma_0 = 1 \ {\rm MPa}$.

This phenomenon is classified as cut-chip-chunk (CCC) wear \cite{CCC1,CCC2,CCC3,CCC4}. The origin lies in the scaling of elastic energy with particle size as $r_0^3$, while fracture energy scales as $r_0^2$. Thus, at sufficiently large scales, there is always more stored energy than needed for particle detachment. The CCC contribution increases with wear particle size up to a cutoff defined by the smallest wavenumber $q_0$ used in the surface characterization, where $(r_0)_{\rm max} = \pi/q_0 = L$ (scan length). In practice, some physical length scale exist
which determined the cut-off length, such as tread block size or the thickness of the rubber layer above the steel belt in slick tires.

\vskip 0.3cm
{\bf 4 Summary and Outlook}

In this review, we have described some of the fundamental ideas and models related to the origin of wear, with a primary focus on rubber wear. We have proposed that the large variation in the wear coefficient $K_0$ observed in the Archard wear law can be naturally explained using the wear model originally proposed by Rabinowicz. This theory predicts a minimum size for wear particles, determined by the elastic and cohesive strength of the material. This cut-off length is typically on the order of $\sim 1 \ {\rm \mu m}$. If smaller wear particles are observed, they are likely produced by a different type of wear process, such as corrosive wear. Another example involves tires operating at high speeds or undergoing abrupt braking, which generates high temperatures and leads to the formation of ultrafine tire wear particles through the condensation of volatilized organic additives and oils\cite{oil1,oil2}. Particle mobility spectrometers have shown that these ultrafine particles exhibit a monomodal size distribution centered at a few tens of nanometers\cite{oil1,oil2,oil3}.

The Rabinowicz wear theory assumes that wear particles form whenever the elastic energy stored in asperity contact regions exceeds the bonding energy required to detach the particles from their surroundings. In the wear theory developed by Persson et al.\cite{RP1,RP2,RP3}, it is assumed that cracks typically propagate only a short distance $\Delta x$ with each asperity contact, unless the stored elastic energy reaches the ultimate tear strength (see Fig. \ref{TearStrength.eps}). Neither the Rabinowicz nor the Persson et al. theories explicitly consider the detailed mechanisms of crack initiation and growth, which are inherently complex, especially when the cracks are very short. Initial discussions on this topic have been provided by Ciavarella\cite{Ciav1,Ciav2}.

An open question is whether the relation $\Delta x(G)$, obtained for macroscopic cracks (e.g., on the order of centimeters), also applies to the short cracks involved in the formation of small wear particles. For macroscopic cracks, the $\Delta x(G)$ relationship depends on complex phenomena occurring in the crack-tip process zone, including cavitation, stringing, viscoelastic energy dissipation, and even bond scission occurring hundreds of micrometers ahead of the crack tip\cite{Creton}. It is plausible that for micrometer-scale cracks, less energy is dissipated, leading to larger values of $\Delta x$ for a given tearing energy amplitude. As a result, cracks may propagate faster at the micrometer scale than expected based on the macroscopic $\Delta x(G)$ relation, an idea supported by the experimental observations in Ref. \cite{RP1}.

These insights underline the need for future studies that bridge the gap between continuum-scale models and the molecular-scale mechanisms underlying wear in soft materials. In particular, understanding how energy dissipation mechanisms vary with length scale and how they influence crack initiation and propagation remains a key challenge. Studies of the nucleation of cracks at surfaces and measurements of the tearing energy relation $\Delta x (G)$ for micrometer-sized cracks at surfaces (which may be possible using optical microscopy) are needed for a better understanding of rubber wear processes. Investigations into the distribution of cracks on rubber surfaces after run-in are also important and may be possible using scanning electron microscopy on bent rubber samples that open up surface cracks. Moreover, the development of multiscale simulation frameworks that integrate atomistic models (e.g., molecular dynamics) with continuum mechanics will be essential to predict wear behavior under realistic conditions. Addressing these challenges will not only deepen our understanding of rubber wear but also support the design of more durable and environmentally friendly materials.

{\bf Acknowledgements:}  
We thank M. Ciavarella for drawing our attention to the beautiful work of Reye. We also thank him and  
J.F. Molinari for comments on the text. This work was supported by the Strategic Priority Research  
Program of the Chinese Academy of Sciences, Grant No. XDB0470200. We thank  
Shandong Linglong Tire Co., Ltd, for support.

\vskip 0.3cm  
{\bf Appendix A}  

Assuming only elastic deformation and neglecting adhesion (see below), for solids with random roughness  
the normalized contact area $A/A_0$ (where $A_0$ is the nominal contact area)  
is a function of $\sigma_0/\xi E^*$, where $\xi$ is the surface root-mean-square  
(rms) slope. When $A/A_0 \lesssim 0.2$,  
then $A/A_0$ depends linearly on the nominal contact pressure $\sigma_0$ and  
$A/A_0 \approx 2 \sigma_0/\xi E^*$ or $A \approx 2 F_0 /\xi E^*$,  
where $F_0 = \sigma_0 A_0$ is the applied normal force \cite{AA1,AA2,AA3,AA4,AA5,AA6,AA7,AA8,AA9}.  
Thus, the contact area $A$ is independent of the nominal contact area  
if $\xi$ is independent of $A_0$. For a surface with a power spectrum with  
a large enough roll-off region, this is always the case. However, if the nominal  
contact area is so small that no roll-off region occurs, then this is not always the case.  
For example, for a self-affine fractal surface with the Hurst exponent $H$  
we have $C(q) \sim q^{-2(1+H)}$ and  
$$\xi^2 = \int d^2q \ q^2 C(q) \sim \int_{q_0}^{q_1} dq \ q^{1-2H}$$ 
$$= {q_1^{2-2H} \over 2(1-H)} \left [ 1- \left ( {q_0 \over q_1} \right)^{2-2H}\right ]\eqno(A1)$$
In general, $q_1 \gg q_0$. For example, if the nominal contact area has a linear  
size of $10^{-2} \ {\rm m}$ and the short  
distance cutoff length is $10^{-9} \ {\rm m}$, we get $q_0/q_1 \approx 10^{-7}$. In this case, if $H<0.9$ we can neglect the term  
$(q_0/q_1)^{2-2H}$ in (A1) and the contact area is, to a good approximation,  
independent of the nominal contact area. However, many surfaces have $H \approx 1$ and, assuming no roll-off,  
for such surfaces the area of real contact will depend  
on the nominal contact area. For example, if $H = 0.95$, then $2-2H = 0.1$ and $(q_0/q_1)^{2-2H} \approx 0.2$.  
However, even in this case, increasing the size of the nominal contact area by a factor of 2 will  
change $(q_0/q_1)^{2-2H}$ from $0.1995$ to $0.2138$, so the contact area changes only by a factor of  
$(1 - 0.1995)/(1 - 0.2138) \approx 1.019$, or by $1.9\%$.

Most surfaces with self-affine fractal-like roughness  
have $H \leq 1$, but there is no theory that excludes $H > 1$, at least in some length-scale interval, and in  
such cases the situation is completely different and the area of real  
contact would depend significantly on the size of the nominal contact region.  
The observation that the friction force is usually independent of $A_0$ indicates that most rough surfaces,  
if they have self-affine fractal-like roughness, have ``normal'' Hurst exponent $H < 1$.

In most cases, if the friction coefficient depends on the normal force $F_0$, this is not due  
to the effect discussed above, but rather because $A/A_0$  
is not small compared to unity, or because adhesion is important. For  
soft solids like silicone rubber (PDMS) with Young's modulus $E \approx 2 \ {\rm MPa}$ and for typical  
surface rms-slope of 1, we get $A/A_0 \approx 0.5$ for $\sigma_0 \approx 0.5 \ {\rm MPa}$, which is a typical  
contact pressure in practical applications.

We have shown that for elastic contact without adhesion, in most cases $A/A_0$ is proportional to the normal force.  
This proportionality holds even when adhesion increases the contact area, assuming that adhesion does not  
manifest itself as a pull-off force, which is a condition that is satisfied in most practical applications, e.g.,  
negligible adhesion when lifting a bottle from a table \cite{AA10}. However, the pressure region over which this proportionality  
holds is reduced by adhesion \cite{AA10}.

\end{document}